\documentclass[12pt,aps,prd,preprint,showkeys,showpacs]{revtex4}
\pdfoutput=1 

\usepackage[utf8x]{inputenc}
\usepackage{amsmath}
\usepackage{amsfonts}
\usepackage{amssymb}
\usepackage{graphicx}

\begin{document}
\title{Neutrino currents in wakes of cosmic strings.}
\author{Sovan Sau and Soma Sanyal}
\affiliation{School of Physics, University of Hyderabad, Gachibowli, Hyderabad, India 500046}


\begin{abstract}

{Neutrinos rotating around Abelian Higgs strings will generate a neutral current close to the string. As the string moves through the cosmic plasma, the velocity 
kick generated by the motion of the string will enhance the neutrino current in the wake region. The neutrino current density depends on its distance from the string and is oscillatory in nature. This leads to neutrino density gradients in 
the plasma. Such a flux of neutrinos with periodic density fluctuations will lead to electron currents in the plasma. The current will act like a cross-perturbation across the cosmic string wake. The perturbation as well as the high Reynolds number of the plasma will result in the generation of magnetic fields in the wake of the cosmic string.}

\end{abstract}
\keywords{cosmic strings, wakes, neutrinos.}

\pacs{25.30.Pt, 98.80.Cq, 47.27.Vf, 95.30.Qd}

\maketitle

\section{Introduction}

Abelian Higgs strings are linear topological defects generated in symmetry breaking phase transitions in the early universe due to the Kibble mechanism \cite{neilsen}. They are often considered the prototypical field theory to study the constraints imposed on the string defects by the recent data from the Cosmic Microwave Background (CMB) 
\cite{CMB2}. Recent studies look at the cosmological effects of these strings in the early universe. Detailed simulations have been performed which show the evolution of these strings \cite{abegrowth}. It is well known that the motion of these cosmic strings generate wakes behind them \cite{wakes}. Various signatures related to cosmic string wakes have also been discussed in the literature. \cite{stringsigna1, stringsigna2}. 

In a recent work, it has been shown that particles that have mass can be trapped close to an Abelian Higgs string \cite{hartman,abhisek}. These particles have a finite angular momentum. The energy, angular momentum and the linear momentum have specific values for which the particles have closed orbits around the Abelian Higgs string. This basically leads to the trapping of the particles close to the string. Now, Abelian Higgs strings once formed will move through the cosmic plasma generating wakes behind them. The plasma around the cosmic string
consists of many particles. The energy of these particles depend on the plasma temperature at that time. We are interested in studying neutrinos in the wakes of Abelian Higgs strings. Since neutrinos have mass, for certain values of angular momentum and energy, they will rotate in closed orbits around a static Abelian Higgs string. We show that due to the inherent lepton asymmetry in the universe, the rotating neutrinos will give rise to a neutrino current close to the string. Generally, cosmic strings are not stationary in
the early universe. So the motion of the string will impart a velocity kick to the neutrinos. The velocity kick imparted by the moving string will enhance the density 
gradients generated by the neutrino current. As the string moves through the plasma, the neutrinos will interact with the background plasma.  Now, non - uniform fluxes of neutrinos have been of interest in the case of supernova explosions, in this work we would like to study non-uniform neutrino currents in the wakes of cosmic strings. These non-uniform neutrino currents are said to generate inhomogeneities in the electron distribution in the background plasma due to the ponderomotive force \cite{bingham,silva}.It has also been predicted that they can generate magnetic fields \cite{shukla}. However, the description of the interaction as a ponderomotive
force brought about criticism and other methods such as kinetic theory was used to look at the same interactions. They concluded that the neutrino gradient could lead 
to instabilities in the background plasma.  Bento \cite{bento} did a  recent study of the neutrino plasma interaction using quantum field theory methods which 
took into account the spin and chiral structure of the weak interaction. He concluded that the previous studies had taken into consideration only the gradient of the electron and neutrino densities, but failed to account for the vector currents. The time derivatives of the vector current also contribute to the dispersion relation.
Though conditions were imposed on the neutrino fluxes generating instabilities in the supernova, all the studies did show that the inhomogeneous neutrino flux would generate an electric current due to the interaction between the neutrinos and the electrons.  An electron current leads to charge separation in the plasma, given the high Reynolds numbers in the early universe, this would mean that magnetic fields would be generated in the wakes of these cosmic strings due to the presence of these neutrino gradients.  
We make an estimate of the magnitude of the magnetic field which can be generated in the wakes of these cosmic string and we find that the estimate is well within the limits of the observational bounds on magnetic fields set by nucleosynthesis calculations.

The question that naturally arises is why are we using the neutrinos around the cosmic string rather than the electrons, which also have mass and will be rotating around the cosmic string. The answer lies in the fact that both electrons and positrons will generate currents of equal magnitudes and in opposite 
directions. So the net effective electromagnetic current is zero. However, in the case of neutrinos, it is well established that a neutrino-antineutrino asymmetry is present in the early universe \cite{leptogenesis}. The order of magnitude of the lepton asymmetry varies from $10^{-10}$ to higher values of $10^{-4}$ depending on the various leptogenesis models. It is this lepton asymmetry in the neutrino sector which will give rise to a neutrino current around the Abelian Higgs strings.   

In section II we first review our understanding of cosmic string wakes. We then calculate the neutrino current around an Abelian Higgs string in section III. We discuss the generation of magnetic fields due to these neutrino currents in the cosmic string wakes in section IV.  In Section V, we present our conclusions. 

\section{Wakes due to cosmic strings}

As is well known, space is locally flat but globally conical around a cosmic string. If the string moves with a velocity $v_s$ in a particular direction in a plane, the particles moving along that plane will get a velocity perturbation $\Delta v$ due to the deficit angle of the string. Since the velocity perturbation will be towards the string, it leads to an overdensity behind the cosmic string which is generally referred to as the cosmic string wake \cite{wakes}. Wake formation has been studied both analytically as well as numerically. Planar wake formation with both hot and cold dark matter have been studied. Since at one time wakes were one of the important candidates for structure formation therefore clustering of baryons in cosmic string wakes has also been studied. We briefly review the clustering of particles  in cosmic string wakes in this section. Our emphasis is on the clustering of neutrinos and electrons in the wake region.

As mentioned before, cosmic string wakes arise due to the conical nature of space time around a cosmic string. A cosmic string has a deficit angle given by $\delta \theta = 8 \pi G \tilde{\mu}$, where $\tilde{\mu}$ is the mass per unit length of the string. As the string moves forward, an observer behind the string would see matter streaming past it. Apart from the velocity of the particles, the particles also feel a velocity kick towards the center of the plane behind the string. The magnitude of the kick is given by $\delta v \sim \delta \theta v_s \gamma_s $ where $v_s$ is the velocity of the string and $\gamma_s$ is the relativistic factor. As more and more particles are kicked towards the string, an overdensity or wake is generated behind the string. A detailed description of wake formation is given in ref. \cite{wakes}. We only mention some of the salient points here which are required for this work. 

Strings generated in the early universe generally move at relativistic velocities. Long strings moving at such high velocities may get chopped into smaller loops. So a long string moving at a time $t_{i}$ will generate a wake whose dimensions are given by $c_1 t_i \times t_i v_s \gamma_s \times \delta \theta  t_i v_s \gamma_s $. Here $c_1$ is a constant of order one \cite{branden21}. The opening angle of the wake depends upon the deficit angle of the cosmic string. 

The overdensity in the wake will lead to further accretion of matter, and in this way the wake will grow in thickness. Generally, the plasma is charge neutral and hence the overdensity too is charge neutral. Though the cosmic strings play a subdominant role in structure formation, the wakes due to the moving strings gives rise to distinct signatures in the background microwave radiation. This is due to the formation of shocks in cosmic string wakes. As the string moves through the plasma the velocity in the direction of motion is greater than the velocity in the other directions. Detailed analysis of string wakes have shown that shocks are generated in the wakes of strings \cite{stebbins}. In relativistic fluid flows, both strong shocks as well as weak shocks can be generated \cite{hiscock}. Shocks formed at high temperatures will have a narrow opening angle and a high overdensity, typically double the background density. As temperature goes down the shocks are more diffuse. Since the density of the plasma is related to the temperature, an overdensity in the shock will mean a temperature gradient in the plasma \cite{layek}. This temperature gradient $\nabla T$ can be calculated at a particular temperature  of the plasma. Since the density is always higher in the shock so the temperature will also be higher there. Now for a general shock structure, the number density gradient and the temperature gradient are parallel to each other. However, this scenario will change for the case of neutrinos moving close to an Abelian Higgs string. Due to the presence of the lepton asymmetry in the early universe, neutrino currents are generated close to the cosmic strings. In the next section, we will first describe the neutrino currents generated near the cosmic string and then in Section III we will describe how these currents create a number density gradient which is not aligned with the temperature gradient. This misalignment between the number density gradient and the temperature gradient will generate a magnetic field in the wake of the cosmic string.

\section{Neutrino current density around cosmic strings}

Recent work on the Abelian Higgs model has shown that particles with mass cluster close to an Abelian Higgs string\cite{abhisek}. Based on the energy and angular momentum of the particles, one can obtain the overdensity around a stationary Abelian Higgs string. Since neutrinos have mass, they will also cluster around the Abelian Higgs string. These particles move around the string in orbits with a finite  angular momentum. Some of them are trapped close to the string core, while others move in escape orbits depending on their energies \cite{hartman}.We will then have a large number of neutrinos rotating around the Abelian Higgs string as it moves through the plasma.  
The Fermi distribution of these neutrinos in a rotating system is given by 

\begin{equation}
f(E, l_z, \beta) = \left [exp\left(\frac{E - l_z \Omega - \mu \beta}{T} \right) + 1     \right]^{-1}
\end{equation}

where $\Omega$ is the angular velocity, $\mu$, is the chemical
potential of the neutrinos, and $l_z$ is the projection of
the particle's total angular momentum  on the direction of $\Omega$ and
 $E$ is the energy of the neutrinos. The factor $\beta$ takes on the values $1$ or $-1$ depending on whether the neutrinos are more than the antineutrinos in the plasma. The Abelian Higgs string is cylindrical in shape so we use the cylindrical coordinates. The most general line element obeying all the symmetry properties pertaining to the Abelian Higgs string is given by 
\begin{equation}
 ds^2 = N^2(r) dt^2 - dr^2 - L^2(r) d\phi^2 - N^2(r) dz^2  
\end{equation}
The factors $L(r)$ and $N(r)$ are determined by the boundary conditions.They are related to the values of the fields at a distance $r$ from the axis of the string (the $z$ axis in this case). Our string moves in the $x- y $ plane and  there is a magnetic field along the $z$-axis whose value depends on the scalar and the vector potentials.
The metric corresponds to a cylindrical metric with a deficit angle $\delta \theta$ as mentioned previously. The deficit angle far from the core of the 
string is proportional to the energy per unit length of the string. 
Further, the Lagrangian of the Abelian Higgs string is usually rescaled and written in terms of two dimensionless constants $  8\pi G \eta^{2}$ and $\frac{\lambda}{e^{2}}$. The deficit angle depends upon these two constants. The cosmic string has a finite width with a core of magnetic flux as well as a scalar core.
The width of these cores are the inverse of the gauge boson mass and the Higgs mass respectively. The geodesics of the neutrinos around the cosmic string are open or closed circular orbits \cite{hartman}. So the neutrino particles appear to rotate about the $z$-axis. To obtain the neutrino current density around the cosmic string, we first need to calculate the appropriate spinor wave function  $\psi(E, p_z , l_z, \beta)$. 

We start with the neutrino field equations in the metric of the cosmic string. 
\begin{equation}
\gamma^{\mu} (\partial_{\mu} - \Gamma_{\mu}) \psi - m \psi = 0 
\end{equation}
Since we are looking at neutrinos with mass, $m$ denotes the mass of the neutrinos. 
A suitable choice of Gamma matrices for this metric are,
$\gamma^{t} = \gamma^{0}$, $\gamma^{r} = \gamma^{1}$,$\gamma^{\phi} = \frac{\gamma^{2}}{L(r)}$,$\gamma^{z} = \gamma^{3}$. Here we have considered $N(r) = 1$ as the particles are considered quite close to the string and their distance from the core is small. 
Using these Gamma matrices, we then solve the field equations by choosing the trial solution in the form,  
\begin{equation}
\chi = e^{-i E t} e^{i p_z z} e^{-i l_z \phi} \begin{bmatrix} \xi \\ \zeta

\end{bmatrix}
\end{equation}
Here $E$ is the energy, $p_z$ in the component of linear momentum in the $z$ direction and $l_z$ is the projection of angular momentum along the $z$-axis. On solving the equations, we get
\begin{equation}
\xi = \begin{bmatrix}
i(E + m - p )^{1/2} J_{l_z+1/2} (\alpha r) \\
\beta (E + m + p )^{1/2} J_{l_z-1/2} (\alpha r)
\end{bmatrix}
\end{equation} 
Here $J_{l_z+1/2} (\alpha r)$ are the Bessel functions. For a system with zero chemical potential $\xi = \zeta$ \cite{vilenkin1}. However, for a non zero chemical potential $\zeta = \beta \xi$. As we have mentioned before, since there is a lepton asymmetry in the early universe and it is manifested in the neutrino sector, we take $\zeta = \beta \xi$.  The wavefunction is then given by, 
\begin{equation}
\psi_{(E p_z l_z \beta)} = \frac{1}{4 \pi} e^{-i E t} e^{i p_z z} e^{-i l_z \phi} \begin{bmatrix} \xi \\ \beta \xi

\end{bmatrix}
\end{equation}
The normalization condition for the wavefunction would be, 
\begin{equation}
\int \psi^{\dagger}_{E p_z l_z \beta} \psi_{E'p'_z l_z' \beta'} r dr d\phi dz = \delta_{l_z,l_z'}\delta_{\beta,\beta'} \delta(p_z - p'_z) \delta (E - E')
\end{equation}
Once the wavefunction is obtained, one can obtain the neutrino current. The direction of the current is along the axis of the string i.e along the $z$-direction, however as the wavefunction depends on the distance from the string, 
the magnitude of the current also varies with the distance from the axis of the string. The current as a function of $r$, (the distance from the string axis) is given by, 
\begin{equation}
j{(E p_z l_z \beta)} = \beta \psi^{\dagger}_{E p_z l_z \beta} \gamma^{t} \gamma^{z} \psi_{E p_z l_z \beta}
\end{equation} 
Hence the total current density as a function of $r$, is given by, 
\begin{equation}
J(\alpha r) = \int_0^{\infty} dE \int^E_{-E} dp \sum_{\beta=\pm 1} \sum_{l_z} f(E,l_z,\beta) j{(E p_z l_z \beta)} (\alpha r) 
\label{current}
\end{equation}
To obtain the neutrino current around the cosmic string we have to solve eqn.\ref{current}. 
Analytically this provides quite a challenge, as it is a summation over Bessel functions. One can obtain an approximate solution close to $r = 0$ but it is the   finite values of $r$ that generate the magnetic field. We need to obtain an estimate of the current numerically since we are going to use it for the calculation of the magnetic field generated later on in the paper. As expected a sum of Bessel functions will generate a sinusoidal curve. We find that the numerical solution is in the form of a sinusoidal curve.

We now explain in detail the numerical values chosen to obtain this graph. The cosmic string is generally characterized by it's symmetry breaking scale. The two cores of the cosmic string have width inverse to the Higgs mass ($M_H$) and the $W$ boson mass ($M_W$). The masses are given by $M_H = \sqrt{\lambda} \eta$ and $M_W = \sqrt{2} e\eta$, where $\eta$ is the symmetry breaking scale. Generally for solving equations numerically all the quantities are made dimensionless. This has been done in previous papers involving particle motion around cosmic strings \cite{hartman, abhisek}.  The symmetry breaking scale is usually used to rescale the variables; so that the momentum $p_z$ is scaled to $p_z/e \eta$, $l_z$ by $l_z/e^2 \eta^2$ and similarly for the energy. To obtain the values in the cosmological scenario, all we have to do is to identify the symmetry breaking scale.  For the numerical values of energy and momentum, we have taken the values from ref.\cite{abhisek}. The values are chosen such that they give rise to clustering of particles around the string. In dimensionless variables, they are $E = 1.083$, $l_z^2 = 0.025$ and $p_z = 0.02$. The symmetry breaking scales for these cosmic strings are very high, usually above the electroweak scales \cite{cue}. The electroweak symmetry breaking scales are of the order of $100$ GeV. If we substitute the value of $\eta$ in the equations, we get the current in terms of $GeV^3$. The numerically obtained values of $J(\alpha r)$ are plotted in figure 1. The distance from the core is in terms of $\alpha r$. Here $\alpha = \sqrt{(E+m)^2 - p_z^2}$. Hence $\alpha$ is in $GeV$. This makes  $\alpha r$ dimensionless. The neutrino current will also depend on the amount of lepton asymmetry in the plasma. However we find that in the range of $10^{-10}$ to $10^{-4}$, there is no significant change in the magnitude of the neutrino current.    

\begin{figure}
\includegraphics[width = 86mm]{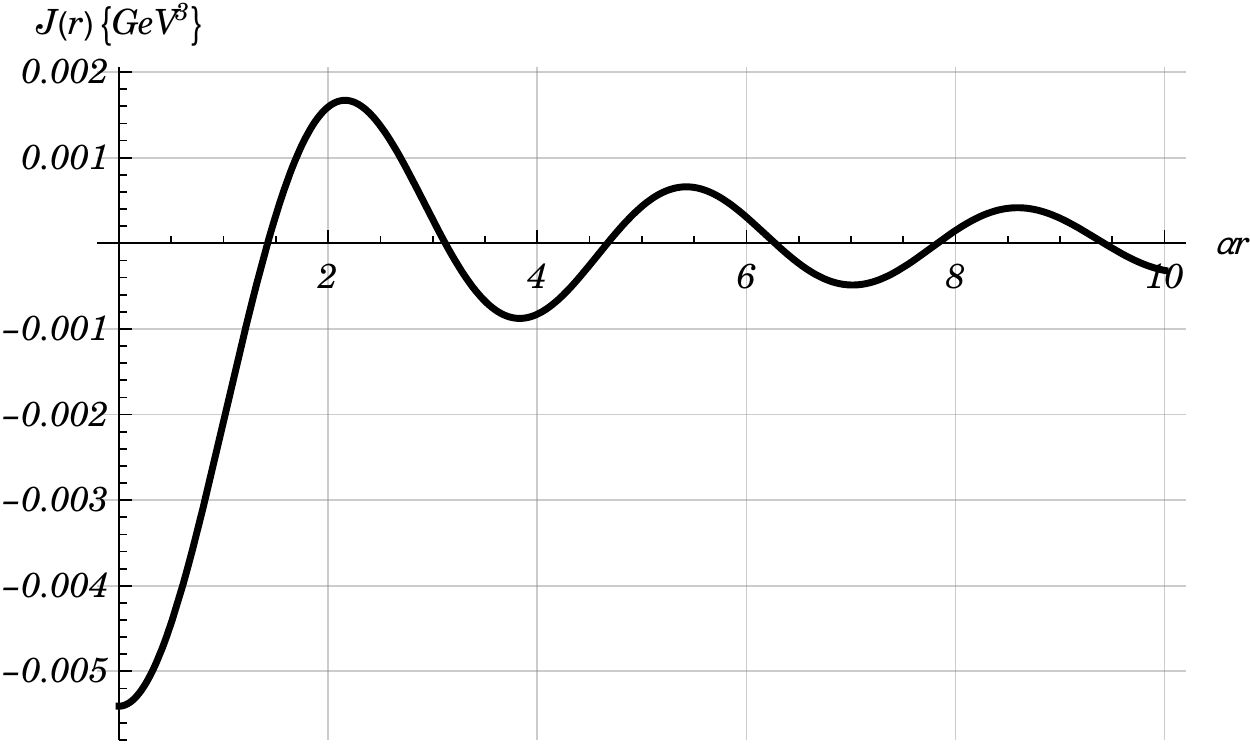}
\caption{Neutrino current as a function of $r$ (distance from the cosmic string). Here $\alpha r$ is dimensionless. }
\end{figure}

The current is oscillatory in nature and decreases with distance from the string. The maximum value is close to the core of the cosmic string. Though the values appear to be small, in the context of the early universe plasma it is not negligible. Detailed studies of neutrino currents in the plasma have been done and  both charged currents and neutral currents have been discussed in the literature \cite{neutrinocurrents}. However, there are no numerical estimates of these currents. These currents depend on the interaction cross section of the particles. 
Unlike these currents we want to emphasize our current is a directed flux of particles moving collectively through the plasma. The numerical estimate given in fig 1  shows that at high temperatures it is not negligible. We will use this estimate when we calculate the magnetic field in the next section.   This solution is for a stationary string; but a cosmic string in the early universe 
is never really stationary. It moves through the plasma with a velocity $v_s$ mentioned before. Behind the string a wake is formed. Now the oscillatory neutrino 
distribution will thus change with time due to the velocity of the moving string. The overdensity behind the string will enhance the current further. As mentioned 
previously, generally the wake density is double the background density of the plasma. So the neutrino distribution will be less in front of the string while it 
will be much more behind the string. Both a spatial as well as a temporal gradient in neutrino density is thus generated in the plasma.

\section{Neutrino currents in moving cosmic string wakes}
As mentioned in the introduction, neutrino interaction with the background plasma has been studied previously. Initial studies by Bingham et. al.\cite{bingham} showed that neutrino gradients in the early universe led to a ponderomotive force in the background plasma. The ponderomotive force was initially a force that occurred in a dielectric in an arbitrary non-uniform electric field. However, it was generalized in \cite{silva} to the interaction of any non uniform field with a
background medium. The non-uniform field was the neutrino field and the background medium were the electrons. The ponderomotive force was due to the weak interaction between the neutrinos and the electrons in a high temperature plasma.  They had shown that a force is exerted on the electrons due to the non-uniform distribution of neutrinos in the plasma. This is the generalized ponderomotive force. The electrons are thus forced to move away from the regions in which the neutrino density is higher.
Considering only the electrons in the plasma, the ponderomotive force was given by \cite{shukla}, 
\begin{equation}
  F_{\nu} = \frac{1}{8 \pi}(f_\nu^2 - 1) \nabla \sum_{k_\nu}\langle | \psi_{k_\nu} |^2 \rangle .  
\end{equation}
Here $f_{\nu}$ is the refractive index of neutrinos in plasma.  However, the ponderomotive force obtained here only contained the terms proportional to the gradients of the neutrino number densities and not the term corresponding to the vector current. Bento \cite{bento} obtained the expression that included the 
terms due to the vector currents using a quantum field theory approach. The other approach was using the relativistic kinetic theory for describing the interaction between the neutrinos and the plasma \cite{silva2}. They showed that for relativistic neutrino jets, plasma instabilities can develop with growth rates of the order of Fermi constant $G_F$. In all these studies, it was 
well established that a electromagnetic current is generated due to the neutrino currents in the early universe. The interaction term in the Lagrangian leads to the force on the electron and both weak-electric and weak-magnetic fields are generated as well as the familiar electromagnetic fields. The electromagnetic fields are given by, \cite{serbeto}
\begin{align}
\vec{E_e} = -\nabla \vec{J_e^0} - \frac{\partial \vec{J_e} }{\partial t}   
&&
\vec{B_e} = \nabla \times \vec{J_e}
\end{align}
The weak-electric and weak magnetic fields are given by, 
\begin{align}
\vec{E_\nu} = -\nabla \vec{J_\nu^0} - \frac{\partial \vec{J_\nu} }{\partial t} 
&& 
\vec{B_\nu} = \nabla \times \vec{J_\nu}
\end{align} 

The flow of neutrinos in a plasma medium is given by the following fluid equations 
\begin{equation}
\frac{\partial N_{\nu}}{\partial t} + \nabla . \vec{J_{\nu}} = 0
\end{equation}
\begin{equation}
\frac{\partial \vec{P_{\nu}}}{\partial t} + ({\vec{v_{\nu}}}. \nabla) \vec{P_{\nu}} = \vec{F_{\nu}} = \sqrt{2} G_F \left( \vec{E_e} + \frac{\vec{v_{\nu}}}{c} \times \vec{B_e} \right)
\end{equation}

Similarly we can also obtain the continuity equations for the electron plasma dynamics, 
\begin{equation}
\frac{\partial N_e}{\partial t} + \nabla . \vec{J_e} = 0
\end{equation}

\begin{equation}
\frac{\partial \vec{P_e}}{\partial t} + ({\vec{v_e}}. \nabla) \vec{P_e} = \vec{F_e} = -e \vec{E_e}+ \sqrt{2} G_F \left( \vec{E_{\nu}} + \frac{\vec{v_e}}{c} \times \vec{B_{\nu}} \right) 
\end{equation}

We then use the standard perturbative approach to obtain the relation between the neutrino perturbation and the electron perturbation in the plasma. We consider $\delta N_e$ to be the electron number density fluctuation and $\delta N_{\nu}$ to the neutrino number density perturbation. The two density perturbations are then related by,
\begin{equation}
\left(\frac{\partial^2}{\partial t^2} + \omega_p^2 \right) \delta N_e = -\frac{\sqrt{2} G_F N_e}{m_e c^2} \left( \frac{\partial^2}{\partial t^2} - c^2 \nabla^2 \right) \delta N_{\nu}
\end{equation}
Here $N_e$ is the mean electron density of the plasma and $\omega _p$ is the plasma oscillation frequency. The neutrino current in the plasma will thus generate a plasma potential $\phi_e$ due to the charge separation given by $\nabla^2 \phi_e = 4 \pi \delta N_e $. An electron current is thus generated in the shock wave.
 
Now we look at our specific case, in our case, behind the cosmic string we have a non - uniform stream of neutrino current. The cosmic string is generated very early in the universe. Once created the string moves through the plasma generating a wake behind it. Apart from the neutrinos the early universe plasma contains a large number density of electrons. Depending on the temperature of the surrounding plasma in the early universe, the other particles may be quark and gluons (for temperatures above the quark - hadron phase transition ) or the neutrons and protons (for temperatures below the quark hadron temperature).
We will look at very high temperatures in the $GeV$ range. The background particles will then be given by the quarks and the leptons. At these temperatures, the particles in the plasma, experience both strongly coupled forces as well as weakly coupled forces.  In such a plasma, perturbative calculations and the quasi-particle description used in deriving the generalized ponderomotive force in the literature \cite{silva,shukla} is not valid. A study by Muller et. al.\cite{muller} has shown that in a high temperature plasma, the shear viscosity of the leptons is dominated by the interaction between the leptons and the quarks. Thus the thermal leptons form a more viscous fluid than the quarks. 
At lower temperatures below the quark hadron transition, the plasma consists of the leptons and the neutrons/protons. The lower temperature plasma is well studied. The velocity of the cosmic string being lower, the plasma can be considered to be non-relativistic. Neutrino plasma interaction in non-relativistic plasmas has been studied in the two fluid hydrodynamic  description. Since both at high temperature and at low temperature, the plasma interaction can be considered as a two fluid hydrodynamic description, we will 
use this to explain how magnetic fields can be generated in the string wakes at any temperature due to the presence of the neutrino density gradients.

As has been established in previous studies the neutrino gradient exerts pressure and results in a local charge separation in the plasma. This leads to a current in the plasma. The current is like a cross-perturbation across the shock. The perturbation is in the form of density gradients in the direction of shock motion. Such perturbations have been studied both numerically and experimentally in the classical regime. There are however no studies for such perturbations for relativistic shocks. Generally if the perturbation is small, the shock remains stable, though the perturbation itself can be accelerated in the direction of the shock. The fact that the leptons in the quark gluon plasma behave as a fluid with a higher viscosity means that accelerated neutrinos may generate  shear induced vorticity in the plasma. A better understanding can only be obtained by a numerical simulation which is beyond the scope of this current work. 

The angle of scattering between the neutrinos and the background plasma determines whether an instability will be generated in the plasma. In this case, there is no head on collision between the neutrino current and the string wake. The angle of scattering is closer to $\pi/2$. A detailed study has shown that for small scattering angle, the elastic process dominates and energy is transferred from the neutrinos to the plasma, however here in this case the angles are closer to $\pi/2$. This means that not much energy is transferred from the neutrinos to the plasma. So no instability is expected to be generated in the plasma. However, the Reynolds number in the plasma is very high. So we have shear induced vorticity in the plasma as well as a high Reynolds number, therefore localized magnetic fields are generated in the string's wake by the Biermann battery mechanism. In a two-fluid description of the plasma with massless electrons, the magnetic field evolution, as given by the Biermann battery mechanism is \cite{biermann}, 
\begin{equation}
\frac{\partial \vec{B_e}}{\partial t} = \nabla \times (\vec{v_e} \times \vec{B_e}) + \frac{\eta_{res}}{4 \pi} \nabla^2 \vec{B_e} - \frac{1}{e N_e } \nabla \times (\vec{j} \times \vec{B_e}) - \frac{1}{N_e e} \nabla N_e \times \nabla T_e
\end{equation} 
Here, $\vec{v_e}$ is the electron fluid velocity, $N_e$ is the number density of the electrons, $T_e$ is the electron temperature and $\eta_{res}$ is the resistivity of the plasma. The last term on the right hand side is the Biermann battery term. In the case where there is no magnetic field ($\vec{B_e} = 0 $), this term generates the magnetic field due to the misalignment between the the density and the temperature gradients of the electrons. We now demonstrate why the density gradient and the temperature gradient would be misaligned due to the neutrino currents in the shocks of the cosmic strings.

\begin{figure}
\includegraphics[width = 86mm]{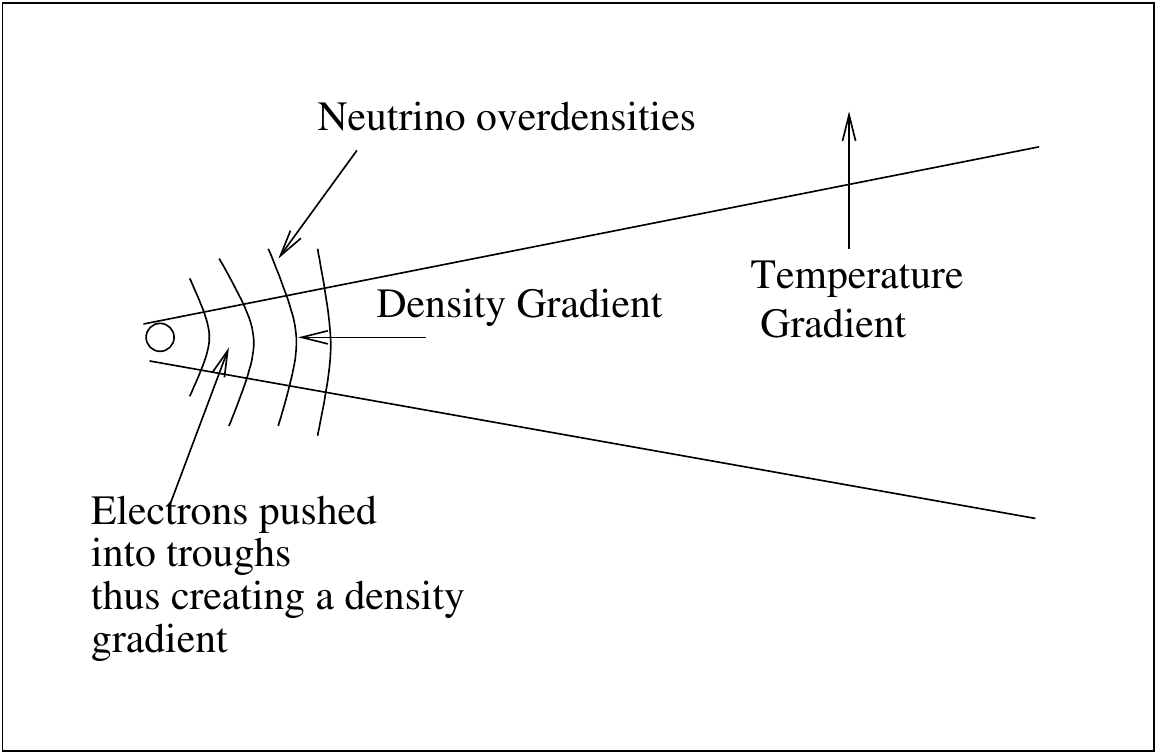}
\caption{An illustration to show the generation of magnetic field in the wakes of cosmic strings due to the non uniform neutrino overdensities. }
\end{figure}

 Let us assume that the string is moving in the $y$ - direction in the $x-y$ plane. The shock is therefore stretched along the $y$ axis with a small width in the $x$-axis. The temperature difference between the overdensity and the background plasma is primarily  along the x-axis. Now, as mentioned before, the neutrino current exerts a force on the electrons which pushes them into regions where there are less neutrinos, so the electron gradients will be complementary to the neutrino gradient caused by the Abelian Higgs strings. This means that if the string was moving in the $y$-direction, they would also be in the $y$-direction. Fig 1. plots the neutrino density with respect to $\alpha r$, where $r$ is the radial distance away from the string, since $r = \sqrt{x^2+y^2}$, the neutrino gradient and the temperature gradient will not be parallel to each other. Since the neutrino gradient and the electron gradient will be complimentary to each other, therefore, the Biermann battery term will give rise to a small but finite magnetic field. 

In fig 2, we have given an illustration to further explain the magnetic field generation. Though an exact calculation is beyond the scope of this work we would like to make an order of magnitude estimate for fields generated at around $100$ GeV temperature scales in the early universe. We assume that the density perturbation for neutrinos is proportional to the density perturbation of the electrons which is taken to be in the $y$ direction. The temperature gradient across a cosmic string shock (in the $x$-direction) is of the order of $10^{-5} T $ \cite{layek2}. So from the Biermann mechanism, 
\begin{equation}
\frac{\partial B}{\partial t} \sim \frac{\sqrt{2} G_F}{m_e c^2} \frac{\partial N_{\nu}}{\partial y} \frac{\partial T_e}{\partial x}
\end{equation} 
We consider $G_F = 10^{-5} GeV^{-2}$, $\frac{\partial N_{\nu}}{\partial y} \sim 0.02 GeV^3$ and 
$\frac{\partial T_e}{\partial x } \sim 10^{-5} \times 200 GeV$. The lengthscales at around the electroweak scales are about $1 GeV^{-1}$ and if we take a similar time scale (about $1 GeV^{-1}$) then the order of magnitude of the generated magnetic field is about $10^{13}$ Gauss. Though this is not a very large field, we point out that this is a conservative estimate as we have not considered the enhancement of the current due to fluid dynamical processes. The $\frac{\partial N_{\nu}}{\partial y}$ is for a stationary cosmic string, as we have mentioned previously, the density inhomogeneities will only be enhanced by a moving string due to the high Reynolds number at those high temperatures. 
At the electroweak scale the equipartition magnetic field is $\sim 10^{24} G$ \cite{baym}. Though the field generated in the shock region is far smaller than this, the Reynolds number is very high at such scale $\sim 10^{12}$, so it is quite possible that this small field can grow into a larger field due to turbulence.  A detailed numerical simulation can give much better estimates of the generated magnetic field.

\section{Summary and Conclusions}

In this work, we have shown that Abelian Higgs strings moving through the plasma generate a neutrino current which has density gradients in the direction of motion of the string. The wake formed behind the string will therefore have a cross perturbation across it. This cross-perturbation creates density discontinuities in the wake. These can be looked as interfaces through which the wakes pass. At very high temperatures, shocks are formed in the wakes of cosmic strings. As the shocks cross the interface, the magnitude of the neutrino current is too small to generate an instability but the particles in the interface are accelerated by the shock wave. Due to the neutrino - electron interaction in the plasma, an electron current is generated in the plasma.
The difference in the viscosity of the neutrino-electron fluid and the neutrino - quark fluid will generate vorticity in the shocks. The Reynolds number of the plasma being high at these temperature, this could lead to the generation of magnetic fields in the shocks of Abelian Higgs strings by the Biermann mechanism. 

At lower temperatures i.e after the quark - hadron phase transition, the plasma consists of the heavier hadrons (neutron/proton) and the lighter leptons. The shock formed behind the string will again have a cross-perturbation. Now, the collision angle between the shock and the neutrinos will be closer to
 $\frac{\pi}{2}$, hence the energy transfer to the plasma will be less. Again no instabilities develop but an electric current is generated which will lead to the generation of a magnetic field. This seems to indicate that the motion of Abelian Higgs strings will always generate accelerated particles and magnetic fields in their wake. 
 
We have obtained an order of magnitude estimate for the generated magnetic field and find that though low it is not negligible. Our estimate is very conservative as it does not include the effect of the high Reynolds numbers in the early universe. There are other mechanisms (such as bubble collisions at the electroweak scale ) which generate lower magnetic fields which are subsequently enhanced due to the high Reynolds number in the plasma. A detailed simulation of the wake structure including the neutrino current will give an idea of the actual magnitudes of the fields generated. That is beyond the scope of our current work. 

In conclusion, neutrino currents around Abelian Higgs string act as a cross perturbation to the wakes generated by the strings. This cross - perturbation leads to the generation of magnetic fields in the wakes of the string at all temperatures. The interaction of a neutrino current with a plasma has been studied before in the context of supernova blasts. There the idea was to generate an instability due to the interaction of the neutrino current and the plasma. In the case of cosmic string, no instability is generated in the shocks at any temperature. Instead, the electron current which is generated by the interaction of the neutrinos with the plasma is of primary importance. The electron current leads to charge separation and the generation of a magnetic field. So Abelian Higgs strings will always generate magnetic fields in the wakes behind them. This can occur at all temperatures in the early universe. We plan to do detailed simulations at a later stage to see if these fields can be the seed magnetic fields that are required to generate the current magnetic fields in the universe.

\begin{center}
 Acknowledgments
\end{center} 
The authors acknowledge discussions with Abhisek Saha and Soumen Nayak. Sovan Sau would like to acknowledge discussions with A.K.Kapoor.


\begin{thebibliography}{100}
 
\bibitem {neilsen}H. B. Nielsen and P. Olesen, 
Nucl. Phys. B 61 (1973) 45; Hindmarsh,M., Stuckey, S. and Bevis, N. Phys Rev {\bf D.79}.123504, 2009 ;

\bibitem{CMB2} J. Lizarraga, et al. 
Journal of Cosmology and Astroparticle Physics 2016.10 (2016): 042.
C. Dvorkin,  M. Wyman, and W. Hu. 
Physical Review D 84.12 (2011): 123519.

\bibitem{abegrowth}D. Daverio, et al.
Physical Review D 93.8 (2016): 085014.

\bibitem{wakes}J. Silk and A. Vilenkin, 
Phys. Rev. Lett.53, 1700 (1984); M. J. Rees,
Mon.Not. R. Astron. Soc.222, 27 (1986); T. Vachaspati, 
Phys.Rev. Lett.57, 1655 (1986); 



\bibitem{stringsigna1} N. Kaiser and A, Stebbins, 
Nature (London) 310, 391 (1984) ;


\bibitem{stringsigna2} O. F. Hernández, Y. Wang, R. Brandenberger, and J. Fong,
J. Cosmol. Astropart. Phys. 08 (2011)014;
E. McDonough and R. H. Brandenberger, 
J. Cosmol. Astropart.Phys. 02 (2013) 045.

\bibitem{hartman}B. Hartmann, and P.J. Sirimachan,  High Energ. Phys. (2010) 2010: 110. 

\bibitem{abhisek} A. Saha and S. Sanyal, Journal of Cosmology and Astroparticle Physics 2018.03 (2018) 022.

\bibitem{cue} Y. Cui, D. E. Morrissey.Phys. Rev.D 79 (8) 083532 (2009). 

\bibitem{neutrinocurrents} V.B. Semikoz and M. Dvornikov, Int. J. Mod. Phys. D 27, 1841008 (2018).

\bibitem{bingham} R. Bingham, J.M. Dawson, J.J. Su, and H.A. Bethe, Phys. Lett.
A 193, 279 (1994); R. Bingham, H.A. Bethe, J.M. Dawson, P.K. Shukla, and J.J. Su, ibid. 220, 107 (1996).

\bibitem{silva} L. O Silva, R. Bingham, J. M . Dawson, W. B. Mori, Phys. Rev. E 59 2273 (1999).

\bibitem{shukla} P.K. Shukla, R. Bingham, J. T. Mendonca and L. Stenflo, Physics of Plasmas, Vol 5, No 7, 2815 (1998).



\bibitem{bento} L. Bento, Phys. Rev. D, Vol 61, 013004 (1999); L. Bento, Phys. Rev. D, Vol 63, 077302 (2001).

\bibitem{serbeto} A. Serbeto, Physics Letters A, 296, 217 (2002) 

\bibitem{leptogenesis} G. Mangano, G. Miele. S. Pastor, O. Pisanti and S. Sarikas, Phys. Letts. B. Vol 708, Issues 1-2, 1-5 (2012) ; R. Buras and D.V. Semikoz, Astropart.Phys.17:245-261, (2002).

\bibitem{branden21} R. H. Brandenberger, R. J. Danos, O. F. Hernández, and
G. P. Holder, 
J.Cosmol. Astropart. Phys. 12 (2010) 028.

\bibitem{stebbins} A. Stebbins, S. Veeraraghavan,
R. H. Brandenberger, J. Silk, and N. Turok, 
Astrophys. J.322, 1 (1987).

\bibitem{hiscock} W.A. Hiscock and J. B. Lail,  Physical Review D 37 869 (1988).

\bibitem{layek}B. Layek, S. Sanyal and A. M. Srivastava, Physical Review D 63, 083512, (2001) 
 

\bibitem{vilenkin1} A. Vilenkin , Physical Review D 20, 1807 (1979)

\bibitem{silva2} L. O. Silva, R. Bingham, J. M. Dawson, J. T. Mendonça, and P. K. Shukla, Physics of Plasmas 7, 2166 (2000)

\bibitem{muller}B. Müller and Di-Lun Yang, Physical Review D 91, 125010 (2015)

\bibitem{biermann} K. M. Schoeffler, N. F. Loureiro, R. A. Fonseca, and L. O. Silva.  Physics of Plasmas 23, no. 5 (2016): 056304.

\bibitem{layek2}B. Layek, Physical Review D. 71(6) 063527, (2005).

\bibitem{baym} G. Baym, D. Bodekar and L. McLerran  Physical Review D 53, 662 (1996)

\end{thebibliography}
\end{document}